\documentclass[aps,prb,twocolumn,showpacs,amsmath,amssymb,superscriptaddress]{revtex4}
\usepackage{dcolumn}
\usepackage{bm}
\usepackage{graphicx}
\usepackage{epstopdf}
\usepackage{color}
\begin{document}

\newcommand{\bfk}{{\bf k}}
\newcommand{\bfR}{{\bf R}}

\title{Chiral $d$-wave superconductivity on the honeycomb lattice close to the Mott state}
\author{Annica M. Black-Schaffer}
 \affiliation{Department of Physics and Astronomy, Uppsala University, Box 516, S-751 20 Uppsala, Sweden}
 \author{Wei Wu}
 \affiliation{D\'{e}partement de Physique and RQMP, Universit\'{e} de Sherbrooke, Sherbrooke, Qu\'{e}bec, Canada}
 \author{Karyn Le Hur}
 \affiliation{Centre de Physique Th\'{e}orique, Ecole Polytechnique, CNRS, 91128 Palaiseau Cedex, France}
\date{\today}

\begin{abstract}
We study superconductivity on the honeycomb lattice close to the Mott state at half filling. Due to the sixfold lattice symmetry and disjoint Fermi surfaces at opposite momenta, we show that several different fully gapped superconducting states naturally exist on the honeycomb lattice, of which the chiral $d+id'$-wave state has previously been shown to appear when superconductivity appears close to the Mott state. Using renormalized mean-field theory to study the $t$-$J$ model and quantum Monte Carlo calculations of the Hubbard-$U$ model we show that the $d+id'$-wave state is the favored superconducting state for a wide range of on-site repulsion $U$, from the intermediate to the strong coupling regime. 
We also investigate the possibility of a mixed chirality $d$-wave state, where the overall chirality cancels. 
We find that a state with $d+id'$-wave symmetry in one valley but $d-id'$-wave symmetry in the other valley is not possible in the $t$-$J$ model without reducing the translational symmetry, due to the zero-momentum and spin-singlet nature of the superconducting order parameter. Moreover, any extended unit cells result either in disjoint Dirac points, which cannot harbor this mixed chirality state, or the two valleys are degenerate at the zone center, where valley hybridization prevents different superconducting condensates.
We also investigate extended unit cells where the overall chirality cancels in real space. For supercells containing up to eight sites, including the Kekul\'{e} distortion, we find no energetically favorable $d$-wave solution with an overall zero chirality within the restriction of the $t$-$J$ model. 
\end{abstract}
\maketitle

%
% -------------------------------------------------- %
% INTRODUCTION
% -------------------------------------------------- %
\section{Introduction}
% Intro to d-wave SC, cuprates:
Superconductivity driven by strong electron-electron interactions has been an enigma in condensed matter physics ever since the high-temperature cuprate superconductors were discovered almost 30 years ago.\cite{Bednorz86} 
Common to all cuprate superconductors are square (or rectangular) CuO$_2$ layers, where superconductivity is known to originate upon doping of the antiferromagnetic Mott insulating phase at half filling.\cite{Anderson87, Harlingen95, Tsuei00RMP,Damascelli03RMP, Tremblay06,Rice12,Gull13} 
The superconducting state has a spin-singlet fourfold $d_{x^2-y^2}$-wave symmetry in the CuO$_2$ planes.\cite{Harlingen95, Tsuei00RMP} The spin-singlet configuration can be viewed as a direct consequence of the antiferromagnetic state in the undoped parent compounds, as the Mott physics present at half filling naturally transfers into a spin-singlet symmetry for the superconducting order appearing at finite doping levels. The $d$-wave symmetry is a result of strong on-site repulsion, which heavily favors order parameters which average to zero over the full Brillouin zone, of which the $d$-wave state is the simplest even-parity possibility. This is in contrast to conventional electron-phonon driven superconductors which have spin-singlet isotropic $s$-wave order parameters.
The spin-singlet $d$-wave symmetry of the superconducting state is thus a direct consequence of strong electron-electron repulsion. Beyond the high-temperature cuprate superconductors,\cite{Taillefer10} many heavy fermion superconductors are likely also $d$-wave superconductors,\cite{Thompson12} as well as some organic superconductors.\cite{Arai01,Shinagawa07,Ichimura08}

% Honeycomb with d-wave, why interesting:
All currently established $d$-wave superconductors have square (or rectangular) lattice symmetry, which naturally hosts a fourfold symmetric order parameter. The sixfold symmetric honeycomb and triangular lattices, on the other hand, offer crystal structures which do not conveniently accommodate a $d$-wave state. As a result, the $d_{x^2-y^2}$- and $d_{xy}$-wave symmetries belong to the same irreducible representation for all hexagonal lattices. This means both solutions necessarily have the same transition temperature $T_c$.\cite{Sigrist91,Black-Schaffer07} Below $T_c$ higher order terms become important in the energy functional and the complex combination $d_{x^2-y^2}\pm id_{xy}$ ($d \pm id'$) is very generally favored.\cite{Sigrist91, BaskaranTria, KumarShastry, HonerkampTria, Ogata03,WangLeeLee,Kuznetsova05, Black-Schaffer07, Nandkishore12,Wu13} The $d \pm id'$ state is a chiral, time-reversal symmetry breaking state with the chirality set by the sign between the two different $d$-wave components\cite{Volovik89, Volovikbook92, Volovik97} and where the non-trivial topology\cite{Schnyder08} guarantees the existence of two co-propagating, i.e., chiral, edge states.\cite{Volovik97,Black-Schaffer12PRL}

Finding a superconductor with a honeycomb or triangular lattice and strong electron-electron interactions thus offers the exciting possibility of realizing chiral $d+id'$ pairing. The honeycomb lattice further has the interesting electronic property that close to half filling the normal-state has not one single Fermi surface but instead two disconnected Fermi surfaces centered around the Brillouin zone corners at $K$ and $K' = -K$. Superconductors with multiple Fermi surfaces, such as the iron-pnictides/chalcogens, are often known to host an intricate gap structure.\cite{Stewart11,Black-Schaffer13multi}
Thus, electron-driven superconductivity close to a Mott insulating phase at half filling on the honeycomb naturally provides a very rich playground for possibly realizing exotic superconducting states.
Beyond superconductivity, Mott physics of the honeycomb lattice has also been proposed to give rise to other exotic states such as topological Mott insulator and exotic quantum spin Hall states, especially in conjunction with finite spin-orbit coupling.\cite{Kane_PRL05_2,Bernevig_2006,Koenig_Science07,Raghu08, Shitade09,Young08, Pesin10, Rachel10, Hohenadler11,Wu12, Cocks12, Reuther12, Ruegg12,LiuLeHur13,Hohenadler13}

% Graphene and sc at vHs:
In terms of materials, the most prominent honeycomb material today is undoubtedly graphene.\cite{Novoselov_science2004, CastroNetoRMP09} 
There already exist a multitude of theoretical proposals for superconductivity driven by electron-electron interactions in graphene.\cite{Black-Schaffer14Chirald} The Fermi surface in undoped graphene only consists of two points, but doping graphene generates more free carriers, which increases the chances of superconductivity. Especially the region around the van Hove singularity at very heavy doping, where the Fermi surface transitions to be centered around $\Gamma$, has been shown to be very promising for superconductivity with $d+id'$-wave symmetry, generated by even weak electron-electron interactions.\cite{Gonzalez08, Nandkishore12, Nandkishore12PRB, Wang11, Kiesel12} 
While this $d+id'$-wave state has been shown to be reasonably stable to disorder,\cite{Lothman14} experimental advancements are still needed in order to reach such high doping levels in graphene.
% SC in honeycomb materials close to half filling.
Superconductivity has also been explored in graphene at lower doping levels where the Fermi surface is centered around $K,K'$, but this requires much stronger electron-electron interactions. 
While the on-site repulsive Hubbard interaction in graphene has been approximated to be as large as half the band width,\cite{Wehling11U} it still only puts graphene in a weak to intermediate coupling regime, where the small low-energy density of states close to half filling effectively prevents superconductivity.
In order to enhance the chiral $d+id'$-wave state in lightly doped graphene to the level of detectability, a proximity effect to external superconductors has been proposed.\cite{Black-Schaffer10, Black-Schaffer13vortex} 
Also bilayer graphene, which has a large density of states at the Fermi level even in the undoped state, has been proposed to host a $d+id'$-wave superconducting state upon weak doping.\cite{Milovanovic12,Roy13,Vafek14, Murray14}

Looking beyond graphene, other honeycomb materials might have more potential in terms of chiral $d$-wave superconductivity close to half filling. One promising material is In$_3$Cu$_2$VO$_9$, where singly occupied $3z^2-r^2$ Cu orbitals form a $S = 1/2$ honeycomb lattice.\cite{Kataev05} Experimentally, the undoped ground state has been identified as a likely N\'{e}el antiferromagnet,\cite{Moller08, Yan12} with $d+ id'$-wave superconductivity proposed to appear upon finite doping.\cite{Wu13,Gu13} Another material is  the (111) bilayer of the perovskite SrIrO$_3$,\cite{Xiao11} which has been found to host an antiferromagnetic Heisenberg coupling on a buckled honeycomb lattice, where finite doping has the possibility to give rise to $d+id'$ superconductivity.\cite{Okamoto13,Okamoto13b} The recently discovered superconductor SrPtAs\cite{Nishikubo11} has also been proposed to have $d+id'$-wave pairing,\cite{Fischer14} although multiple bands around the Fermi level add additional complexity to the electronic structure\cite{Goryo12,Youn12} not present in simple one-orbital honeycomb lattice systems.
Another interesting materials family is the iridates A$_2$IrO$_3$ (A = Na, Li), where the Ir atoms form a honeycomb lattice and magnetism has been shown to be captured by a Heisenberg-Kitaev model.\cite{Kitaev06,Jackeli09, Chaloupka10,Singh12,Witczak14,Reuther14} Hole doping this model for dominant antiferromagnetic Heisenberg coupling has been shown to give spin-singlet $d+id'$ pairing, whereas spin-triplet $p+ip'$-wave phases dominate when the Kitaev coupling is increased.\cite{Hyart12, You12, Scherer14}
Similar physics could potentially also be implemented using ultracold atoms, which have recently been used to produce a tunable honeycomb lattice.\cite{Tarruell12}

% Close to half filling theory for SC:
Theoretically, methods ranging from mean-field theory,\cite{Black-Schaffer07, Wu13} to functional renormalization group,\cite{Honerkamp08,Wu13} quantum Monte Carlo (QMC),\cite{Pathak10, Ma11} and Grassman tensor-product state variational methods,\cite{Gu13} have already shown that a chiral $d+id'$-wave state can appear on the honeycomb lattice close to half filling in models with strong electron-electron interactions. Predominantly the Hubbard model with moderate on-site repulsion,\cite{Ma11} and especially the $t$-$J$ model,\cite{Black-Schaffer07, Honerkamp08, Pathak10, Wu13, Gu13} which is the resulting model in the limit of very strong Hubbard on-site repulsion,\cite{Hirsch85, GrosJoyntRice87, Choy95} have been used. 
For example, renormalized mean-field theory (RMFT)\cite{ZhangGros88, Anderson04, Edegger07, Ogata08, LeHur09} and a slave-boson\cite{ReadNewns83, Coleman84, Kotliar86, Affleck88, LeeRMP06}  approach have recently been used to handle the no-double occupancy criterion in the $t$-$J$ model.\cite{Wu13} The chiral $d+id'$-wave state was shown to emerge quite generally upon doping, in agreement with earlier mean-field studies not incorporating the no-double occupancy criterion.\cite{Black-Schaffer07} Further, it has also been suggested that the disjoint Fermi surfaces might give rise to a state with $d+id'$ symmetry in one valley but $d-id'$ symmetry in the other valley.\cite{Tran11, Wu13} In this {\it mixed chirality state} the chirality cancels out leaving the system time-reversal symmetry invariant and with no edge states, but still with a highly unconventional order parameter.

% Our work:
In this work our goal is to study the competition between single versus the mixed and other zero net sum chirality $d$-wave superconductivity on the honeycomb lattice when doping the Mott state at half filling.
To accomplish this we first of all provide a general symmetry analysis of the allowed superconducting symmetries on the honeycomb lattice resulting from strong electron-electron interactions, especially focusing on the nodal structure close to half filling. Since the normal-state Fermi surface is disjoint and centered around $K,K'$ close to half filling, its nodal structure can be notably different from that of a single Fermi surface centered around the zone center $\Gamma$.
We then specialize and study superconductivity from strong on-site repulsion. More specifically, we use RMFT to study the $t$-$J$ model and QMC calculations to treat the original Hubbard-$U$ model. The RMFT results are specifically focused on the quasiparticle energy spectrum and provide evidence that the mixed chirality state is not more energetically favorable than the single chirality state. 
Our QMC study complements earlier QMC work in the intermediate-$U$ regime,\cite{Ma11} as it spans a whole range of $U$ values also approaching the strong-$U$ limit, and shows that the $d+id'$ state gets more favorable at higher on-site $U$ values, which corroborates the use of RMFT.
Using RMFT we then provide a detailed analysis of the proposed mixed chirality state.\cite{Tran11, Wu13} By using the general symmetry analysis we conclude that the mixed chirality state cannot be a physically viable superconducting state of the $t$-$J$ model, or other models with spin-singlet superconductivity, without reducing translational symmetry.
Moreover, even with reduced translational symmetry the Dirac points are either still disjoint, in which case they cannot still harbor the mixed chirality state, or they are degenerate at the zone center, where valley hybridization prevents different superconducting condensates.
Following this result we also study real-space modulations in four-, six-, and eight-site unit cells. These extended unit cells allow the order parameter phase winding (or flux) to be zero or alternating on neighboring honeycomb plaquettes, which can result in a zero net sum chirality in the material. The six-site cell also allows for a Kekul\'{e} distortion which has no identifiable plaquette flux.\cite{Roy10,Roy13}
In all cases we find that the single chirality $d+id'$ state is energetically favorable. By extrapolating these results we draw the conclusion that real-space modulations of the chiral $d$-wave state, which generate a net zero sum chirality state, do not lead to a lower energy. 
We can thus conclude that the single chirality $d$-wave state is very likely the most stable state on the honeycomb lattice for superconductivity close to the Mott phase at half filling.

% Structure:
The structure of the article is as follows. In Section \ref{sec:symmetry} we provide a general symmetry analysis for zero-momentum pairing on the honeycomb lattice, as appropriate for pairing originating from strong electron-electron interactions. In Section \ref{sec:tJ} we study the $t$-$J$ model within RMFT, paying special attention to the quasiparticle energy, and in Section \ref{sec:QMC} we provide a QMC study of the Hubbard model which corroborates the RMFT findings.
Then in Section \ref{sec:extcells} we use a symmetry analysis to show that a mixed chirality state is not feasible without breaking translation symmetry. We also study extended unit cells and conclude that no real-space modulation leads to a mixed or net zero sum chirality state being the favored state in the $t$-$J$ model. Finally, in Section \ref{sec:conclusions} we provide a concluding discussion and summary.

% -------------------------------------------------- %
% GENERAL SYMMETRY ANALYSIS:
% -------------------------------------------------- %
\section{General symmetry analysis}
\label{sec:symmetry}
We start with a general symmetry analysis of the superconducting state close to half filling on the honeycomb lattice.
The superconducting state always breaks the global U(1) symmetry, but can also break additional symmetries present in the normal state Hamiltonian, such as crystal, spin-rotation symmetries but also time-reversal symmetry. The term ``unconventional superconductivity" is often used to classify states with such additionally broken symmetries. Even without detailed knowledge of the attractive interaction causing superconductivity it is possible to do a general symmetry analysis of the possible superconducting states for a specific material. This is due to the fact that the quadratic BCS Hamiltonian describing the superconducting state in momentum space results in a linear eigenvalue equation for the superconducting order parameter $\Delta$ around the transition temperature $T_c$. We can therefore expand the momentum space dependence of $\Delta$ with respect to a set of basis functions, which can be classified according to the irreducible representations of the symmetry group of the normal-state Hamiltonian (see, e.g., Refs.~[\onlinecite{Sigrist91,Black-Schaffer14Chirald}] for a more detailed treatment). Apart from accidental degeneracies, the superconducting state will belong to a single irreducible representation at $T_c$ and it is only possible to have a mixture of symmetries belonging to different irreducible representation if multiple superconducting transitions take place as the temperature is lowered. 

The two-dimensional (2D) honeycomb lattice belongs to the hexagonal group $D_{6h}$ with $k_z = 0$, which restricts the irreducible representations to those even under in-plane reflection. We will here limit ourselves to zero-momentum pairing, which pairs electrons symmetrically across the $\Gamma$ point in the Brillouin zone, and then $D_{6h}$ is the relevant symmetry group for all doping levels of the honeycomb lattice. The pairing from strong electron-electron interactions is very short range in real space and thus extended in $\bfk$-space, which naturally results in zero-momentum pairing close to half filling on the honeycomb lattice.
We will also for the moment only consider a single band, but as we shall see in Section \ref{sec:tJ}, extensions to multiple bands can be straightforward.
Furthermore, we restrict the treatment to spin-singlet and $s_z = 0$ spin-triplet solutions in spin-space, since all unitary $s_z = \pm 1$ states can be constructed by spin-rotation of the $s_z = 0$ spin-triplet solution.

Table \ref{tab:sym} shows the simplest possible basis functions satisfying the symmetry requirements of the  irreducible representations of the $D_{6h}$ group for the 2D honeycomb lattice. Higher order expansions are of course possible but give order parameters with more zero-energy nodes. Since the quasiparticle energy is $E_\bfk = \sqrt{\varepsilon_\bfk^2 + |\Delta_\bfk|^2}$ for a single band with normal-state dispersion $\varepsilon_\bfk$ and a unitary superconducting order parameter $\Delta_\bfk$, states with lower number of nodes are usually energetically favored.
%
% TABLE:
\begin{table}
\begin{tabular}{ccl}
\hline \hline
 Irreps. & Basis functions & Brillouin zone symmetry \\
 \hline
 $A_{1g}$ & 1, $k_x^2 + k_y^2$ & 
 \parbox[c]{0em}{\includegraphics[height =2.2cm]{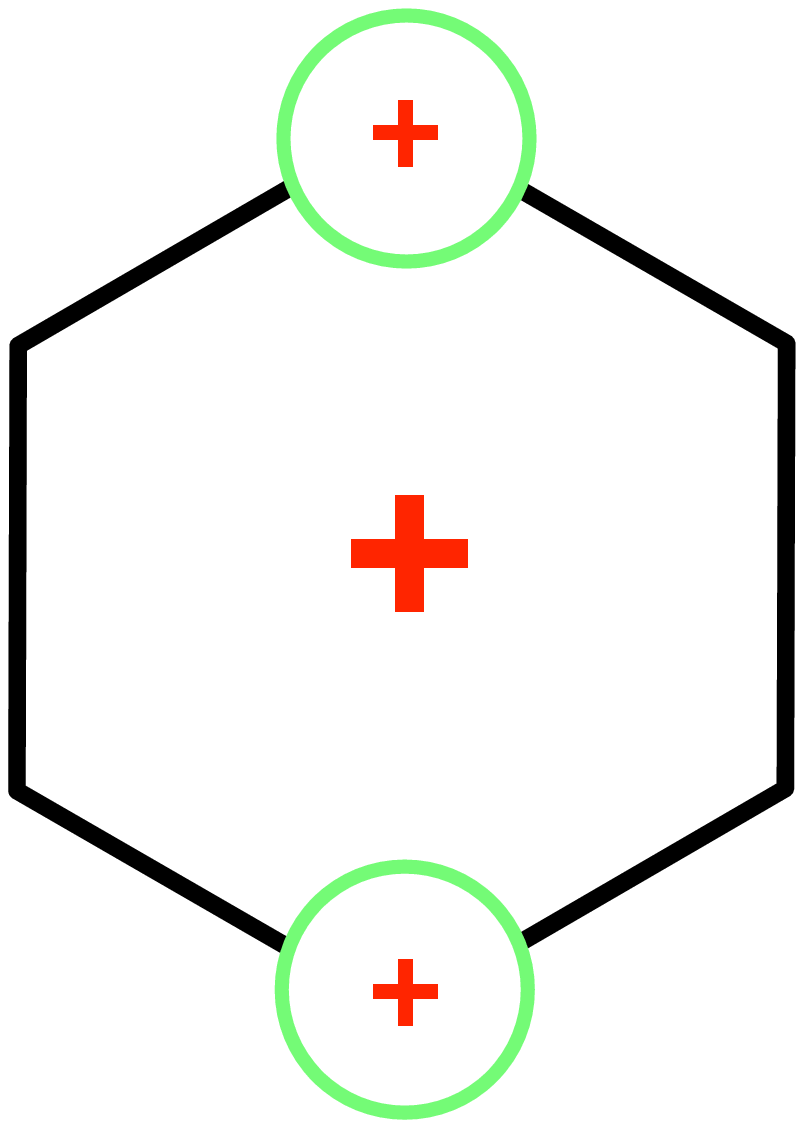} }\\
 \hline
 $A_{2g}$ & $k_x k_y (k_x^2 - 3k_y^2)(k_y^2 - 3k_x^2)$ & 
 \parbox[c]{0em}{\includegraphics[height =2.2cm]{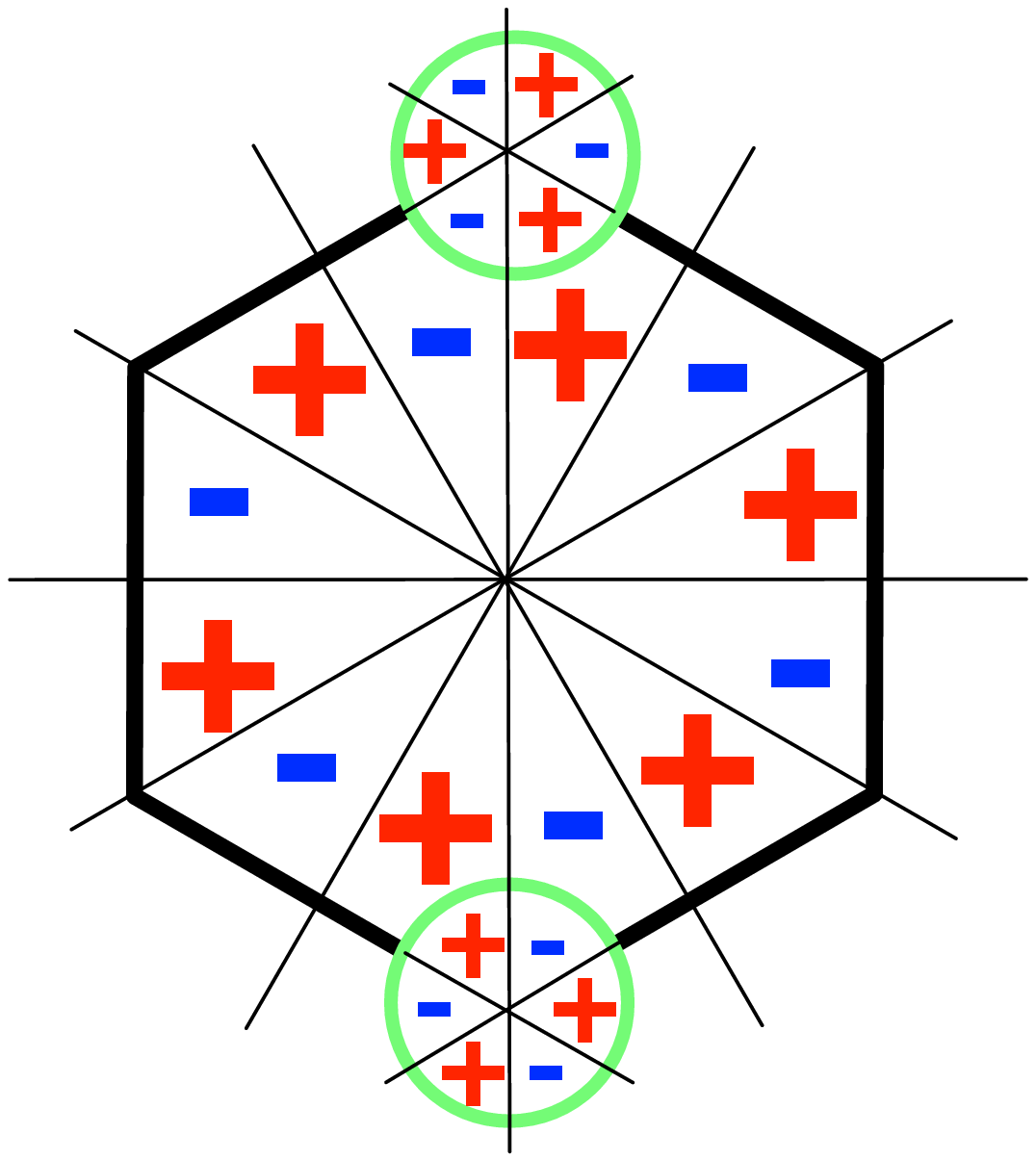} }\\
 \hline
 $E_{2g}$ & $(k_x^2-k_y^2, 2k_x k_y)$ & 
 \parbox[c]{0em}{\includegraphics[height =2.2cm]{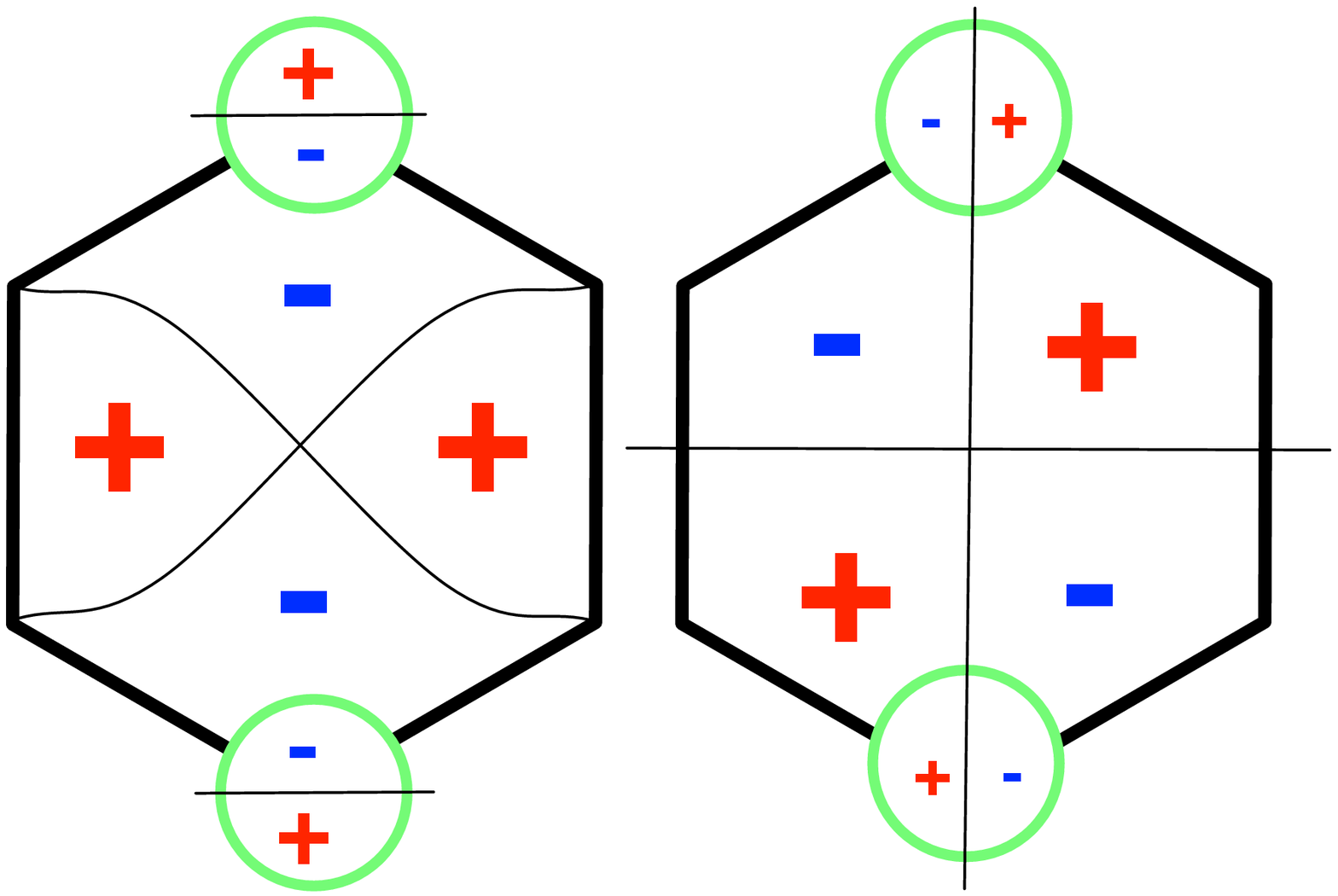} }\\
  \hline
$B_{1u}$ & $k_x(k_x^2-3k_y^2)$ & 
 \parbox[c]{0em}{\includegraphics[height =2.2cm]{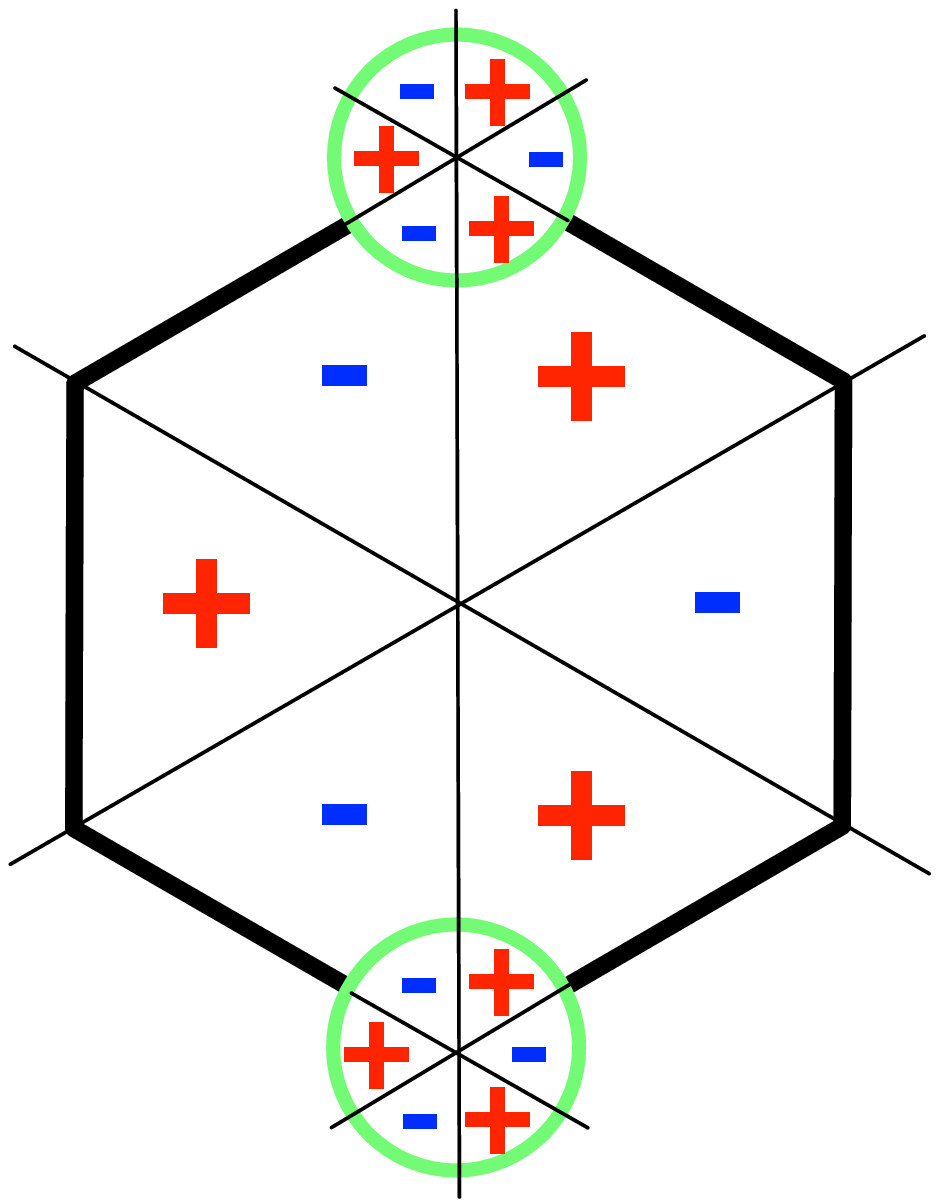} }\\
  \hline
$B_{2u}$ & $k_y(k_y^2-3k_x^2)$ & 
 \parbox[c]{0em}{\includegraphics[height =2.2cm]{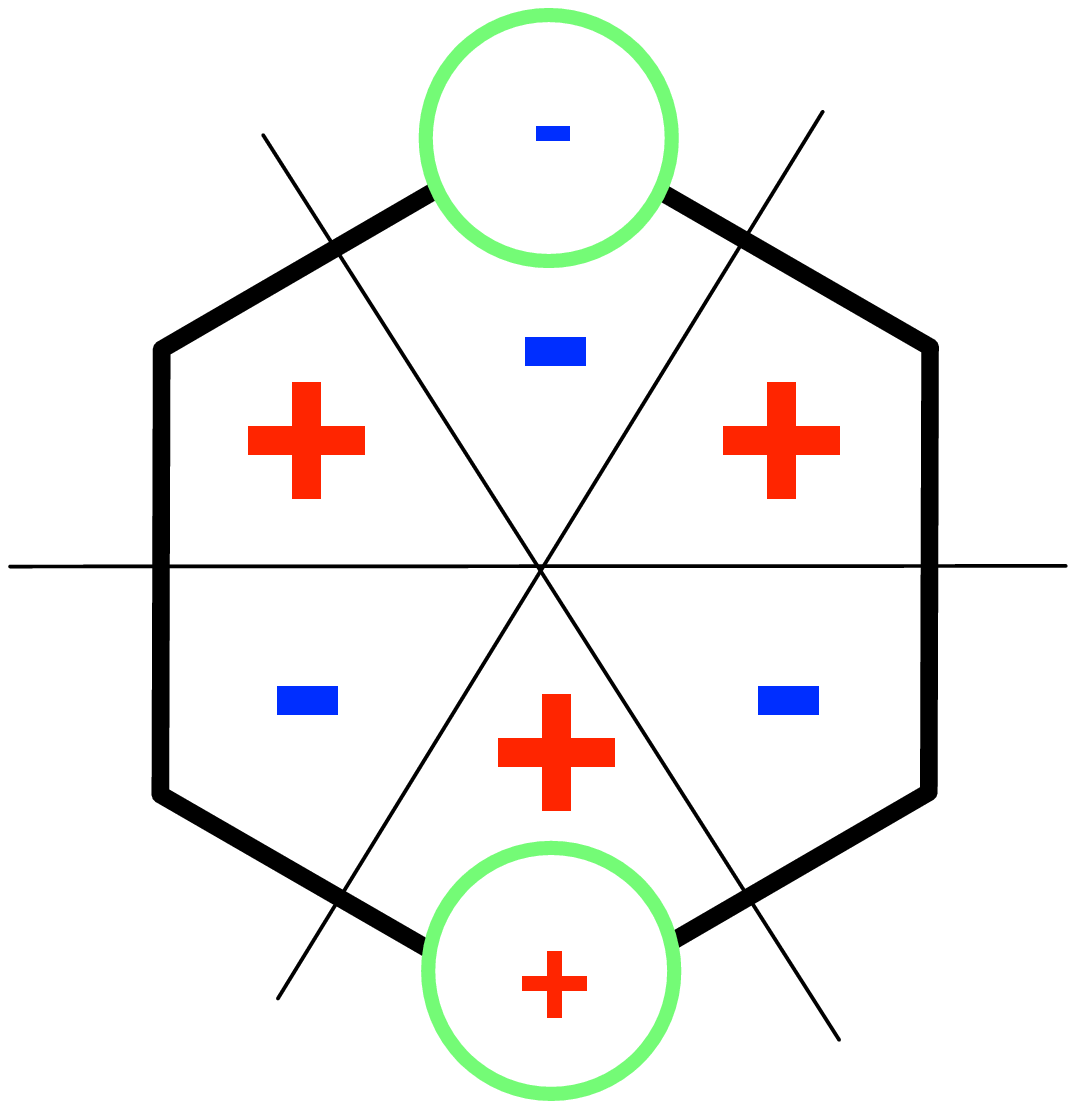} }\\
  \hline
$E_{1u}$ & $(k_x, k_y)$ & 
 \parbox[c]{0em}{\includegraphics[height =2.2cm]{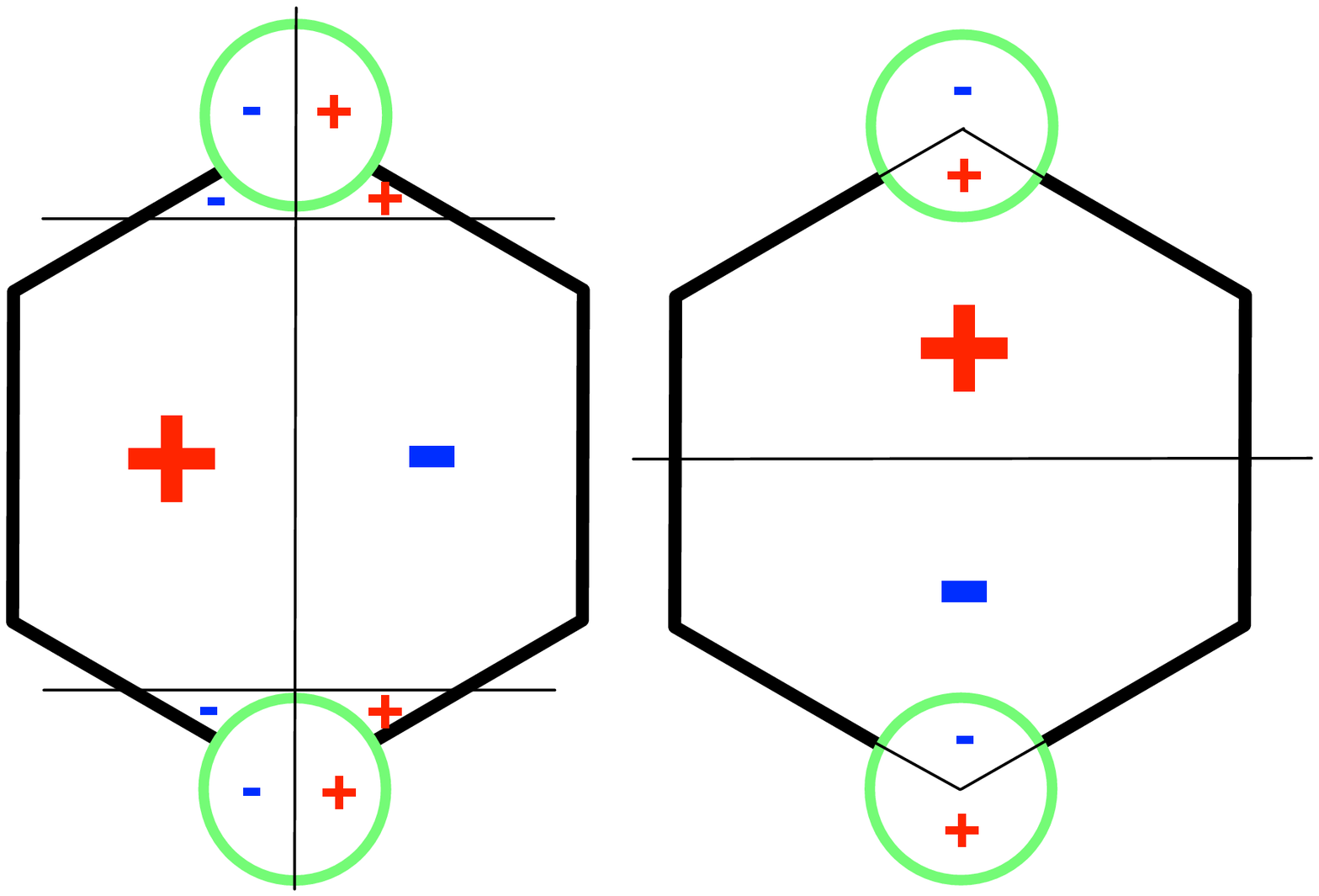} }\\
  \hline \hline
 \end{tabular}
  \caption{Available irreducible representations (irreps.) for the honeycomb lattice ($D_{6h}$ crystal group with $k_z = 0$) with their simplest basis functions and symmetries in the hexagonal Brillouin zone (thick black lines). Green circles indicate the Fermi surface close to half filling, which is disconnected and centered around $K$ and $K'$. For filling beyond the van Hove singularity at $\pm 1/4$ doping the Fermi surface is instead centered around the zone center at $\Gamma$.
}
 \label{tab:sym}
\end{table}
The basis functions should be understood as representatives of the transformation behavior but they might not obey the translation symmetry of the particular lattice. Taking also the latter into account results in the last column in Table \ref{tab:sym}, which schematically shows the simplest (in the sense of lowest number of nodes around $\Gamma$ and $K,K'$) symmetries for each irreducible representation. As seen, some of the basis functions with two- and fourfold rotational symmetry needs to be slightly modified in order to also obey translational symmetry in reciprocal space. 
Since the superconducting order parameter is necessarily fermionic in nature, even-parity ($g$) representations correspond to spin-singlet states, whereas spin-triplet pairing gives odd-parity ($u$) spatial dependence for a single band. 
 
%
% A1g and A2g:
\subsection{$A_{1g}$ and $A_{2g}$ irreducible representations}
A superconducting state belonging to the identity representation $A_{1g}$ has all the symmetries of the normal-state Hamiltonian. This is possible either for a constant order parameter, or if it has the same $\bfk$-dependence as the normal-state Hamiltonian. The former is the conventional $s$-wave superconducting state, typically present for phonon-driven superconductors, whereas the latter is an extended $s$-wave state. The extended $s$-wave state has an effective $p_x+ip_y$ dependence around each Dirac point and has been suggested as a possible state for nearest-neighbor attraction.\cite{Uchoa07,Bergman09} It is also usually present as a less favorable solution for interactions which favor a higher angular momentum spin-singlet pairing, as is the case of the $t$-$J$ model close to half filling.\cite{Black-Schaffer07,Wu13} The $s$-wave symmetries give a fully gapped superconducting state, with the exception of exactly at half filling for the extended $s$-wave solution. 

%\subsection{$A_{2g}$ irreducible representation}
The other spin-singlet one-dimensional (1D) representation is $A_{2g}$ has six nodal lines crossing in the zone center. Even though the number of nodal lines are reduced to three around $K,K'$, the high number of nodes makes this state highly unlikely in a real material.

%
% E2g:
\subsection{$E_{2g}$ irreducible representation}
Due to the incompatibility of the sixfold symmetry of the honeycomb lattice with the fourfold symmetric $d$-wave solutions, they belong to a 2D irreducible representation, $E_{2g}$. This means that the two $d$-wave solutions, or any linear combination of them, are necessarily degenerate at $T_c$. Below $T_c$, higher order corrections become important, and the chiral $d\pm id'$-wave symmetries have been shown to be favored quite generally.\cite{Kuznetsova05, Black-Schaffer07, Nandkishore12, Wu13} This follows from a very general Ginzburg-Landau argument where a $\pi/2$ phase shift between two separate order parameters usually lowers the total energy, see, e.g.,~Ref.~[\onlinecite{Sigrist91}]. Due to the intrinsically complex order parameter, these chiral states also break time-reversal symmetry.
While the $d$-wave solutions have two nodal lines crossing a normal-state Fermi surface centered around $\Gamma$, only one nodal line crosses small Fermi surfaces centered around $K,K'$. The $d_{x^2-y^2}$-wave state thus has effectively $p_y$-wave symmetry locally around the $K,K'$ points, whereas the $d_{xy}$-wave state has local $p_x$ symmetry.\cite{Linder09b} Note that this is still compatible with spin-singlet pairing since the order parameter is even over the full Brillouin zone. While the individual $d$-wave solutions will always be nodal superconductors, the chiral $d \pm id'$-wave combinations are fully gapped, apart from the pathological situation when the normal-state Fermi surface only consists of points at $K,K'$ or $\Gamma$. This by itself in fact provides a simple energy argument for why the chiral $d\pm id'$ combinations are preferred.
The chiral $d$-wave superconducting state close to half filling has been shown to be the favored state for both the $t$-$J$ model and the Hubbard-$U$ model for strong interactions.\cite{Black-Schaffer07, Honerkamp08, Pathak10, Ma11, Wu13, Gu13} Moreover, a $d+id'$-wave state has also been shown to be the preferred symmetry in the weak-coupling limit of the Hubbard model when the normal-state Fermi surface is centered around $\Gamma$ \cite{Raghu10, Nandkishore14} and also close to the van Hove singularity.\cite{Gonzalez08, Nandkishore12, Wang11, Kiesel12}

% B1u and B2u:
\subsection{$B_{1u}$ and $B_{2u}$ irreducible representations}
The allowed 1D spin-triplet order parameters belong to the $B_{1u,2u}$ irreducible representations. Both of these have three nodal lines intersecting at $\Gamma$ and are thus both spin-triplet $f$-wave states with very similar properties when the normal-state Fermi surface is centered around $\Gamma$. However, their nodal lines are rotated $30^\circ$ relative to each other, such that the $B_{1u}$ has nodal lines also through $K,K'$, but $B_{2u}$ does not. This makes for distinctly different behavior close to half filling. While the $B_{1u}$ symmetry still has $f$-wave symmetry around $K, K'$, the $B_{2u}$ symmetry gives a fully gapped state with an effective $s_{\pm}$-wave symmetry; i.e., the gap is (largely) constant on each Fermi surface but with different signs. Note that the $s_{\pm}$ state is necessarily a spin-triplet state on the honeycomb lattice since the two Fermi surfaces are located at opposite momenta.
This spin-triplet $s_\pm$-wave state has been found to be the ground state for the Hubbard model on the honeycomb lattice close to half filling in the limit of weak interactions.\cite{Raghu10, Nandkishore14} It has also been shown to be favored in superconductors with generally disconnected, reasonably well-nested, Fermi surfaces driven by a spin-fluctuation exchange mechanism.\cite{Kuroki01}

\subsection{$E_{1u}$ irreducible representation}
The spin-triplet 2D representation $E_{1u}$ consists of the $p_x$-, and $p_y$-wave symmetries, which both have one nodal line across the Brillouin zone. Even after adapting these symmetries to the honeycomb lattice, it is possible to only have one nodal line around $K,K'$. These states are thus essentially equivalent to the spin-singlet $d$-wave solutions close to half filling, apart from having different spin quantum numbers. Similarly to the spin-singlet $d+id'$-wave solution, the $p_x+ip_y$ ($p+ip'$) combination should have the lowest energy below $T_c$ and is also a fully gapped state. 
The $E_{1u}$ solution appears, for example, as the ground state for models with pairing on nearest-neighbor bonds which is odd in $\bfk$. Such spatial odd-parity occurs for spin-triplet bond pairing, but can also be a spin-singlet state if there is an additional quantum number under which the order parameter is odd. A prototype example of the latter is band index, with spin-singlet, $p+ip'$-wave odd-interband pairing found in the $t$-$J$ model,\cite{Black-Schaffer07} as briefly discussed in Section \ref{sec:tJ}. Allowing also for $s_z = \pm 1$ pairing, the spin-triplet $p$-wave states found in the Heisenberg-Kitaev model with dominant Kitaev coupling belong to this irreducible representation.\cite{Hyart12,You12, Scherer14}

To summarize the results of Table \ref{tab:sym}, the honeycomb lattice allows for superconducting order parameters that have many different number of nodes in the Brillouin zone but for disjoint normal-state Fermi surfaces at $K,K'$ the number of nodes present in the superconducting state is often reduced. 
Most interestingly, there exist several different fully gapped states close to half filling, even for order parameters that change signs in the Brillouin zone such that they average to zero. This is especially important for superconductivity driven by strong electron-electron interactions, which want to avoid on-site pairing. These states include the spin-singlet $d\pm id'$ states belonging to the $E_{2g}$ irreducible representation and the spin-triplet $f$-wave state belonging to the $B_{2u}$ representation. Both of these states have been shown to appear in models of electron-driven superconductivity on the honeycomb lattice close to half filling: the $d+id'$-wave state in models involving strong interactions which favor a spin-singlet configuration \cite{Black-Schaffer07, Honerkamp08, Pathak10, Ma11, Wu13,Gu13} and the $f$-wave state for weak interactions where a spin-triplet state is allowed.\cite{Raghu10, Nandkishore14}
Also the spin-triplet $p$-wave states in the $E_{1u}$ irreducible representation can be fully gapped and have been shown to appear in the Heisenberg-Kitaev model appropriate for some iridate compounds.\cite{Hyart12, You12, Scherer14}

% -------------------------------------------------- %
% RMFT OF THE T-J MODEL
% -------------------------------------------------- %
\section{RMFT on the $t$-$J$ model}
\label{sec:tJ}
Having reviewed the possible order parameter symmetries on the honeycomb lattice in the last section, we now turn to using explicit models for capturing the effect of strong interactions in honeycomb lattice materials, such as In$_3$Cu$_2$VO$_9$, and determining the favorable superconducting symmetries. We will first study the $t$-$J$ model within renormalized mean-field theory (RMFT):\cite{ZhangGros88, Anderson04, Edegger07, Ogata08, LeHur09}
%
% EQUATION:
\begin{align}
\label{eq:HtJ}
H_{tJ}  = & -t g_t \! \! \sum_{\langle i,j\rangle,\sigma} \! \! (a^\dagger_{i\sigma}b_{j\sigma} + {\rm H.c.}) + \mu \sum_{i,\sigma} (a^\dagger_{i\sigma}a_{i\sigma} + b^\dagger_{i\sigma}b_{i\sigma}) \nonumber \\
& + J g_J \sum_{\langle i,j \rangle} ({\bf S}_{i} \cdot {\bf S}_j - \frac{1}{4}n_i n_j).
\end{align}
Here $a_{i\sigma}$ ($b_{i\sigma})$ is the annihilation operator on the A (B) site of the honeycomb lattice for site index $i$ and spin $\sigma$ with $n_i$ being the number operator, $\langle i,j\rangle$ denotes the summation over nearest neighbors, $t$ is the nearest-neighbor hopping amplitude, $\mu$ is the chemical potential measuring doping away from half filling, ${\bf S}_i$ is the spin operator on site $i$, and $J$ is the effective coupling constant.

The $t$-$J$ model can be derived from the Hubbard-$U$ model:
%
% EQUATION:
\begin{align}
\label{eq:HU}
H_{U}  = & -t \! \! \sum_{\langle i,j\rangle,\sigma} \! \! (a^\dagger_{i\sigma}b_{j\sigma} + {\rm H.c.}) + \mu \sum_{i,\sigma} (a^\dagger_{i\sigma}a_{i\sigma} + b^\dagger_{i\sigma}b_{i\sigma}) \nonumber \\
& + U \sum_i (a_{i\uparrow}^\dagger a_{i\uparrow}a_{i\downarrow}^\dagger a_{i\downarrow} + b_{i\uparrow}^\dagger b_{i\uparrow}b_{i\downarrow}^\dagger b_{i\downarrow}),
\end{align}
in the limit of very strong interactions, $U\gg t$. In perturbation theory $J = 4t^2/U$ to lowest order when at the same time double-site occupancy is prohibited through a Gutzwiller projector.\cite{Gutzwiller63, Gutzwiller65, Brinkman70,Hirsch85, GrosJoyntRice87, Choy95} The latter is implemented in Eq.~(\ref{eq:HtJ}) in an average way by the use of the statistical weighting factors $g_t = 2\delta/(1+\delta)$ and $g_J = 4/(1+\delta)^2$, where $\delta = 1-n$ is the doping level. \cite{Vollhardt84, ZhangGros88} 
Exactly at half filling Eq.~(\ref{eq:HtJ}) results in an ordered magnetic state since $g_t  = 0$, which agrees with experimental results for In$_3$Cu$_2$VO$_9$. Increasing the doping away from half filling gives a finite kinetic energy and there is then a possibility for a spin-singlet superconducting state to appear.\cite{Anderson73, Anderson87}

The use of statistical weighting factors and further treating the interaction part in Eq.~(\ref{eq:HtJ}) within mean-field theory is usually referred to as RMFT or Gutzwiller resonance valence bond (RVB) theory of the $t$-$J$ model.\cite{ZhangGros88, Anderson04, Edegger07, Ogata08, LeHur09}
The benefit of a RMFT treatment of the $t$-$J$ model is that it provides straightforward access to the physics and order parameter symmetries of the strongly interacting limit in the Hubbard model. In Section \ref{sec:QMC} we show that the RMFT results are in agreement with QMC results of the original Hubbard model. 
A mean-field solution of the $t$-$J$ model on the honeycomb lattice, resulting in $d+id'$-wave superconductivity, was derived in Ref.~[\onlinecite{Black-Schaffer07}], but there no site occupancy limitations were considered. More recently, a proper RMFT treatment has also been performed.\cite{Wu13} Here we present important additional information on the quasiparticle excitation spectrum of the $d+id'$ superconducting state, but for self-containment we still include a complete derivation of the results. 

To decompose the interaction in Eq.~(\ref{eq:HtJ}) in mean-field theory, we follow the same procedure as for the square lattice, \cite{ZhangGros88,BrinckmannLee01,LeeRMP06,Edegger07} using the mean-field order parameters:
%
% EQUATION:
\begin{align}
\label{eq:OPdef}
\chi_{ij} &= \frac{3}{4}g_J J \sum_\sigma \langle a_{i\sigma}^\dagger b_{j\sigma}\rangle \nonumber \\
\Delta_{ij} & = \frac{3}{4}g_J J \langle a_{i\downarrow} b_{j\uparrow} - a_{i\uparrow} b_{j\downarrow}\rangle.
\end{align}
The $\chi$-field renormalizes the kinetic energy and we will therefore assume that it is uniform in space. The superconducting order is spin-singlet bond pairing. We will for now assume that it is translational invariant in the bulk but allow for variations on the three nearest-neighbor bonds of the honeycomb lattice. This results in $\Delta_{ij} = \Delta_\alpha$, where $\alpha = 1,2,3$ label the three different bonds, with $\bfR_\alpha$ the corresponding bond vector. With these mean-field order parameters we arrive at a quadratic mean-field Hamiltonian:\cite{Black-Schaffer07, Wu13}
%
% EQUATION:
\begin{align}
\label{eq:HMF}
H_{\rm MF}  = & -(t g_t+\frac{1}{2}\chi) \sum_{\langle i,j\rangle,\sigma} a^\dagger_{i\sigma}b_{j\sigma} + {\rm H.c.}  \nonumber \\
+ & \tilde{\mu} \sum_{i,\sigma} (a^\dagger_{i\sigma}a_{i\sigma} + b^\dagger_{i\sigma}b_{i\sigma}) \nonumber \\
 - & \frac{1}{2}\sum_{i, \alpha} \Delta_\alpha (a_{i\uparrow}^\dagger b_{i+\bfR_\alpha\downarrow}^\dagger - a_{i\downarrow}^\dagger b_{i+\bfR_\alpha\uparrow}^\dagger) + {\rm H.c.}  %\nonumber \\
%+ & \frac{3N |\chi|^2}{2Jg_J} + \frac{N\sum_\alpha |\Delta_\alpha|}{2Jg_J}.
\end{align}
Here we have introduced an effective chemical potential, since in the end we determine $\tilde{\mu}$ self-consistently by fixing  $\delta$. We have also ignored constant terms which are only important when calculating the total free energy.

To proceed we first Fourier-transform Eq.~(\ref{eq:HMF}) and then use the transformation $a_{\bfk\sigma} = 1/\sqrt{2}(c_{\bfk \sigma} + d_{\bfk \sigma})$, $b_{\bf\sigma} = \exp(-i\varphi_\bfk)/\sqrt{2}(c_{\bfk \sigma} - d_{\bfk \sigma})$, where $\varphi_\bfk = {\rm arg}(\sum_\alpha e^{i\bfk \cdot \bfR_\alpha})$, to write the Hamiltonian in a basis where the kinetic energy is diagonal, i.e., in the band basis:
%
% EQUATION:
\begin{align}
\label{eq:Hk}
H_{\rm MF} = & \sum_{\bfk \sigma} (\varepsilon_\bfk +\tilde{\mu}) c_{\bfk\sigma}^\dagger c_{\bfk \sigma} + (-\varepsilon_\bfk +\tilde{\mu}) d_{\bfk\sigma}^\dagger d_{\bfk \sigma} \nonumber \\
- & \sum_\bfk \Delta_\bfk^i (c_{\bfk \uparrow}^\dagger c_{-\bfk \downarrow}^\dagger - d_{\bfk \uparrow}^\dagger d_{-\bfk \downarrow}^\dagger ) \nonumber \\
- & \sum_\bfk \Delta_\bfk^I(d_{\bfk \uparrow}^\dagger c_{-\bfk \downarrow}^\dagger - c_{\bfk \uparrow}^\dagger d_{-\bfk \downarrow}^\dagger ).
\end{align}
Here we have introduced the intraband pairing:
%
% EQUATION:
\begin{align}
\label{eq:Di}
\Delta_\bfk^i = \frac{1}{2}\sum_\alpha \Delta_\alpha \cos(\bfk \cdot \bfR_\alpha - \varphi_\bfk),
\end{align}
and the interband pairing:
%
% EQUATION:
\begin{align}
\label{eq:DI}
\Delta_\bfk^I = \frac{i}{2}\sum_\alpha \Delta_\alpha \sin(\bfk \cdot \bfR_\alpha - \varphi_\bfk),
\end{align}
as well as the band energy $\varepsilon_\bfk = -\tilde{t} \left|\sum_\alpha \exp(i\bfk \cdot \bfR_\alpha)\right|$, with $\tilde{t} = (tg_t +\chi/2)$. 
In terms of discussing energetics it is clearly favorable to work in the band basis as we here have direct access to the normal-state Fermi surface ($\varepsilon_\bfk  = \pm \tilde{\mu}$). It is important to note that in the atomic basis used in Eq.~(\ref{eq:HMF}) the kinetic energy is not diagonal and thus the order parameter and its nodal structure in $\bfk$-space cannot alone give any useful information on the quasiparticle spectrum. On the other hand, the drawback of the band basis in Eq.~(\ref{eq:Hk}) is that there is now both intraband and interband gaps. However, away from half filling the interband pairing is significantly smaller as it is not on-shell energy wise, and can to a first approximation be ignored.\cite{Black-Schaffer07}

\subsection{Order parameter symmetries}
There are three linearly independent self-consistent solutions to Eq.~(\ref{eq:Hk}) as found in Ref.~[\onlinecite{Black-Schaffer07}]. The simplest solution is a uniform bond solution $\Delta_\alpha = \Delta_0$, which leaves the interband pairing zero. The intraband pairing has a $\bfk$-dependence through Eq.~(\ref{eq:Di}), which is that of the band structure $|\varepsilon_\bfk|$. Thus this is an extended $s$-wave solution belonging to the $A_{1g}$ irreducible representation. The remaining two solutions are degenerate at $T_c$ with the solution space spanned by $\Delta_{d} \sim (2,-1,-1)$ and $\Delta_{d'} \sim (0,1,-1)$ on the three nearest-neighbor bonds. The intraband pairing has for these two solutions the $d_{x^2-y^2}$ and $d_{xy}$ symmetries of the $E_{2g}$ irreducible representation in Table \ref{tab:sym}. The interband pairing has $p_x$ and $p_y$ symmetry and belongs to $E_{1u}$. However, note that the interband pairing is still a spin-singlet state as it is odd in band index.
As discussed in the previous section, below $T_c$ the time-reversal symmetry breaking chiral combinations $d\pm id'$ for the intraband pairing (and $p\pm ip'$ for interband pairing) are very generally favored for $E_{2g}$ pairing.\cite{Black-Schaffer07} The two chiral $d$-wave solutions have the bond order parameters $\Delta_{d\pm id'} \sim (1,\exp(\pm i2\pi/3),\exp(\pm i4\pi/3))$, which are just complex linear combinations of $\Delta_d$ and $\Delta_{d'}$. Note that in the translationally invariant bulk, the two chiral $d$-wave solutions are degenerate at all temperatures.
We will hereafter refer to this 2D solution space as the $d+ id'$-wave states since intraband pairing is generally dominant. 

\begin{figure}[htb!]
\includegraphics[scale = 0.87]{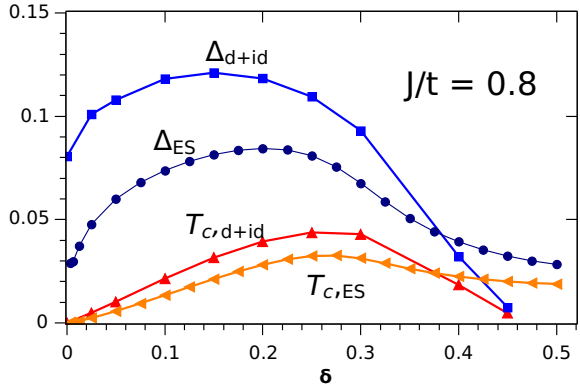}
\caption{\label{fig:phase} (Color online). The superconducting order parameter $\Delta_0$ in units of $\frac{3}{4}g_J J$ for $d+id'$-wave and extended $s$-wave (ES) symmetries as a function of the doping level $\delta = 1-n$ for $J/t = 0.8$.
Within RMFT $T_c$ can be approximated by the quantity $g_t \Delta_0$ which is also plotted.
}
\end{figure}

Numerically we can solve the self-consistency equations for the superconducting bond order at fixed doping levels and calculate the size of the order parameters. The detailed approach for this has already been presented in Ref.~[\onlinecite{Wu13}] and, since this part is not the main goal here, we will not repeat the details but simply present the $T = 0$ phase diagram in Fig.~\ref{fig:phase}.
Superconductivity forms a typical dome structure as a function of doping, and chiral $d$-wave pairing is clearly the favored symmetry state over a wide doping range extending from half filling and past the van Hove singularity at $\delta = 1/4$, where the Fermi surface transitions from being centered around $K,K'$ to being centered around the zone center $\Gamma$. Only for doping far beyond the van Hove singularity does the extended $s$-wave solution become the favorable superconducting solution, in agreement with earlier mean-field theory results.\cite{Wu13,Lothman14}
We also plot $g_t\Delta_0$ which is an approximation of $T_c$ within RMFT.\cite{ZhangGros88, Anderson04, Edegger07, Ogata08} As seen, $T_c$ is zero at half filling, in agreement with having a magnetic state, but then rises as the doping level increases.

\subsection{Quasiparticle spectrum}
\begin{figure}[thb]
\includegraphics[scale = 0.39]{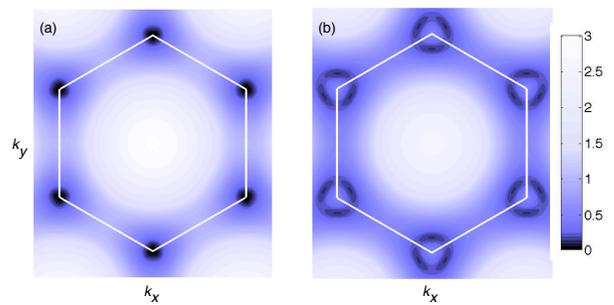}
\caption{\label{fig:EQP} (Color online). Lowest positive quasiparticle energies for the illustrative cases of $\tilde{t} = 1$, $\Delta_0 = 0.4\tilde{t}$, and $\tilde{\mu} = 0$ (a) and $\tilde{\mu} = 0.5\tilde{t}$ (b) for the $d \pm id'$-wave superconducting states. The color scale is chosen such that only $E_{QP} = 0$ is black, which is only present in (a). In (b) there is always a finite energy gap.
}
\end{figure}

To gain more understanding of the favored $d+id'$-wave solution we analytically investigate its quasiparticle spectrum. Due to complex order parameters $\Delta_\alpha$ and both intra- and interband pairing, care has to be taken when diagonalizing the Hamiltonian (\ref{eq:Hk}) and simplifying the expression for the quasiparticle energy in order to arrive at
%
% EQUATION:
\begin{widetext}
\begin{align}
\label{eq:EQP}
E_\bfk = \pm \sqrt{\varepsilon_\bfk^2 + \tilde{\mu}^2 + |\Delta^i_\bfk|^2 + |\Delta^I_\bfk|^2 \pm \sqrt{4\varepsilon_\bfk^2\tilde{\mu}^2 + 2|\Delta^I_\bfk|^2(2\varepsilon_\bfk^2 + |\Delta^i_\bfk|^2) + (\Delta^i_\bfk)^2 (\Delta^I_\bfk)^{\dagger 2} + (\Delta^I_\bfk)^2(\Delta^i_\bfk)^{\dagger 2}}}.
\end{align}
\end{widetext}
The last two terms are different from Ref.~[\onlinecite{Wu13}], where real-valued intra- and interband pairings were used.  However, note that this difference only slightly modifies the self-consistent values of $\Delta_0$ (see Fig.~\ref{fig:phase}) and has no significant influence on $T_c$. In Fig.~\ref{fig:EQP} we plot the lowest positive branch of the quasiparticle energy for chiral $d+id'$-wave pairing for both the half-filled case $\tilde{\mu} = 0$ and for doping away from half filling, in order to determine the lowest energy excitations in the superconducting system. The quasiparticle spectrum is identical for the other chiral solution, $d-id'$.

As seen, the quasiparticle spectrum is fully gapped for doping levels away from half filling. At half filling there exist only nodal quasiparticles at $K,K'$, but this situation does not correspond to a superconducting state since then $g_t = 0$.  This spectrum is to be expected from a $d+id'$-wave symmetric superconducting state in a one-band model, as illustrated in Table \ref{tab:sym}, and the results shows that the interband pairing is not important for the quasiparticle spectrum. In fact, artificially setting the interband pairing to zero generates quasiparticle plots qualitatively similar to those in Fig.~\ref{fig:EQP}. This further justifies calling this state a $d+id'$ state, which technically only refers to the intraband pairing symmetry. We also see how the initially fourfold symmetric $d$-wave solutions adapt to the sixfold honeycomb lattice by giving the $d+id'$ solution (finite) quasiparticle energy minima at three points around each Fermi surface, thus naturally incorporating the lattice symmetry. 
%
%
% FIGURE:

Before closing this section we briefly also comment on the effect of further-neighbor coupling. Expanding the Hubbard Hamiltonian up to fourth-order processes generates a coupling $J_2$ acting on next-nearest neighbors.\cite{Clark11,MacDonald88} Introducing a small such term, as relevant for, e.g., In$_3$Cu$_2$VO$_9$,\cite{Liu13} preserves the antiferromagnetic order at half filling\cite{Clark11} and has been shown to further enhance the $d+id'$ solution over the extended $s$-wave state close to half filling.\cite{Wu13} 

% -------------------------------------------------- %
% QMC OF THE HUBBARD MODEL
% -------------------------------------------------- %
\section{QMC on the Hubbard model}
\label{sec:QMC}
The above RMFT results shows that for strong on-site electron repulsion, approximated by the $t$-$J$ model, the $d+id'$-wave superconducting state is favored for doping levels ranging from half filling to beyond the van Hove singularity. Here we present QMC results of the original Hubbard-$U$ model, which provides additional evidence of the preference for chiral $d+id'$-wave superconductivity in this full doping regime. Since these are the results of true many-body calculations they are highly complementary to the RMFT solution, which relies on first performing a perturbative expansion in $U/t$ and then using mean-field theory. 

More specifically, we perform determinant quantum Monte Carlo (DQMC)\cite{bss1981} calculations on a 24-site cluster for several different values of the Hubbard-$U$ term (here we set $t = 1$ for simplicity) for a range of filling fractions $n$.
We calculate the pairing susceptibilities in the various channels via
\begin{align}
\chi =\frac{1}{N}\times\frac{1}{G_{l}}\int_{0}^{\beta}\sum_{i',i} & \langle\Delta(i',\tau)\Delta^{\dagger}(i,0)\rangle d\tau,
\end{align}
where $N$ is the number of cluster sites and $G_{l}$ denotes the renormalization factor of the pairing form factors: $G_{l}  =  \sum_{\alpha}|g_{l,\alpha}|^{2}$.
The $g_{l,\alpha}$ represent the form factor associated with pairing on bond $\alpha$ in real space, subject to the different irreducible representations of the $D_{6h}$ point group, enumerated by index $l$. For the spin-singlet and spin-triplet pairing amplitudes we have the following pairing possibilities:
% EQUATION:
\begin{align}
\begin{split}
\Delta_{l}^{\dagger}(i) & =\sum_{\alpha}\frac{g_{l,\alpha}^{\dagger}}{\sqrt{2}}(c_{\uparrow i}^{\dagger}c_{\downarrow i+\bfR_\alpha}^{\dagger}-c_{\downarrow i}^{\dagger}c_{\uparrow i+\bfR_\alpha}^{\dagger}),\quad S=0\\
\Delta_{l}^{\dagger}(i) & =\sum_{\alpha}\frac{g_{l,\alpha}^{\dagger}}{\sqrt{2}}(c_{\uparrow i}^{\dagger}c_{\downarrow i+\bfR_\alpha}^{\dagger}+c_{\downarrow i}^{\dagger}c_{\uparrow i+\bfR_\alpha}^{\dagger}),\, S=1,s_{z}=0\\
\Delta_{l}^{\dagger}(i) & =\sum_{\alpha}g_{l,\alpha}^{\dagger}c_{\uparrow i}^{\dagger}c_{\uparrow i+\bfR_\alpha}^{\dagger},\quad S=1,s_{z}=1\\
\Delta_{l}^{\dagger}(i) & =\sum_{\alpha}g_{l,\alpha}^{\dagger}c_{\downarrow i}^{\dagger}c_{\downarrow i+\bfR_\alpha}^{\dagger},\quad S=1,s_{z}=-1 .
\end{split}
\end{align}

In order to extract useful information of the many-body effects on the pairing in the case of small cluster calculations it is usually more appropriate to investigate the connected pairing susceptibility $\chi - \chi_0$ rather than the $\chi$ itself.\cite{SWhite1989} This is due to the fact that $\chi_0$ is the purely disconnected counterpart of $\chi$ (''bubble contribution'') and it contains the Hubbard-$U$ effect on a single-particle level. 

\begin{figure*}[htb]
\includegraphics[scale = 0.35]{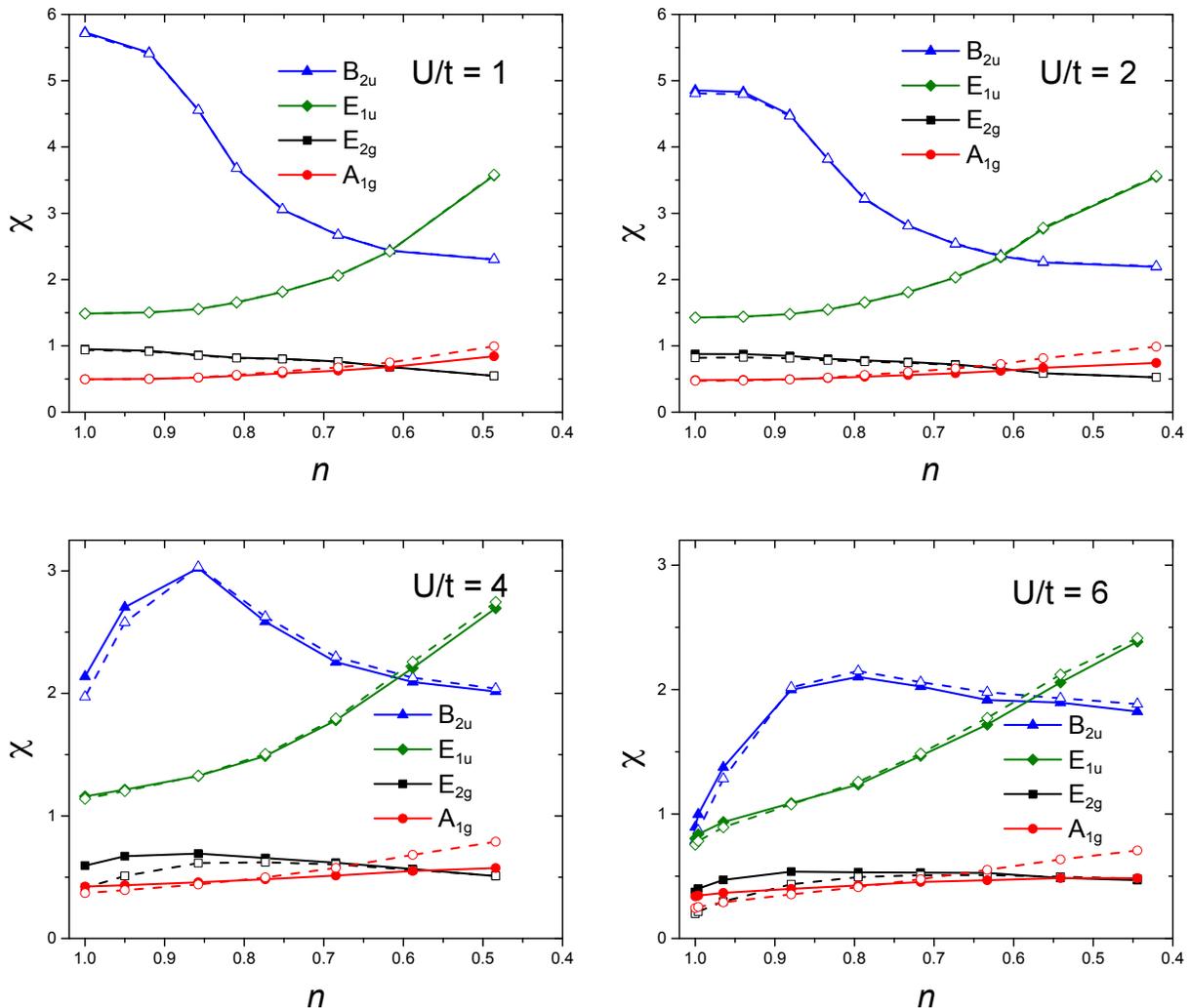}
\caption{\label{fig:QMC} (Color online). Pairing susceptibility $\chi$ (solid lines) and its bubble contribution $\chi_0$ (dashed lines) as a function of electron filling $n$ for four different values of the on-site Hubbard $U/t = 1,2,4,6$ and calculated for the temperature $T/t = 0.2$. The connected pairing susceptibility $\chi - \chi_0$ reveals the impact of the many-body effects from $U$, whether to enhance ($\chi - \chi_0> 0$) or suppress ($\chi - \chi_0< 0$) pair formation. Error bars are smaller than the symbols.
}
\end{figure*}

Thus we plot in Fig.~\ref{fig:QMC} both $\chi$ (solid lines) and $\chi_0$ (dashed lines) as functions of filling $n$, of which the relative magnitude reflects whether pairing is promoted ($\chi > \chi_0$) or suppressed ($\chi < \chi_0$)
by correlation effects. We here include all pairing symmetries with one or fewer nodal lines close to the Brillouin zone corners at $K,K'$.
It is clearly seen in Fig.~\ref{fig:QMC} that for small $U=1,2$, the difference $\chi - \chi_0$ 
is always close to zero, regardless of pairing symmetry and filling $n$. This suggests that the vertex
corrections do not allow pair formation at small $U$ values. At larger $U = 4, 6$ we start seeing clear positive $\chi - \chi_0$ differences for several symmetries, indicating that the pairing with these particular symmetries is enhanced. 
Although the $B_{2u}$, $E_{2u}$, and $A_{1g}$ symmetries all show some positive difference $\chi - \chi_0$  at certain filling fractions $n$ for these larger $U$ values, 
we find that $d_{x^2-y^2}$ (and $d_{xy}$) always has the largest difference $\chi - \chi_0$ for filling fractions between half filling and the van Hove singularity. This shows that it is the most dominant pairing on honeycomb lattice close to half filling. Note that the band width is $6t$, so these QMC results do not only confirm the results of the RMFT calculations for the strong-$U$ limit, but also show that the chiral $d+id'$-wave state is the dominant superconducting instability even in the intermediate-coupling regime.

% FIGURE:
%

% -------------------------------------------------- %
% MIXED CHIRALITIES
% -------------------------------------------------- %
\section{Mixed chirality $d$-wave states}
\label{sec:extcells}
The chiral $d$-wave state found to be the ground state in both the RMFT and QMC calculations above has a twofold ground-state degeneracy,  $d+id'$ or $d-id'$, at all temperatures below $T_c$.  The two states have different chiralities, which can be classified by a Chern number or a Skyrmion winding number,\cite{Volovik89, Volovikbook92, Volovik97} which takes values $\mathcal{N} = \pm 2$ for the two different chiralities. The most notable consequence of the finite topological number is that it guarantees the existence of two chiral edge states crossing the bulk energy gap.\cite{Volovik97,Black-Schaffer12PRL} Beyond the different chiralities, the two $d$-wave states are otherwise identical; for example they have the exactly same quasiparticle spectrum, as seen in Eq.~(\ref{eq:EQP}). 

Most notably, both chiral $d$-wave states break time-reversal symmetry. However, the disconnected Fermi surfaces on the nearly half-filled honeycomb lattice has been proposed to favor $d+id'$-wave symmetry in the $K$ valley but $d-id'$-wave symmetry in the $K'$ valley.\cite{Tran11,Wu13} This {\it mixed chirality} state does not break global time-reversal symmetry and has canceling edge states. In fact, if one plots the gap order parameters in the original atomic basis, the $d+id'$ state is gapped at $K$ but gapless at $K'$, whereas the $d-id'$ state is gapped at $K'$ but gapless at $K$.\cite{Wu13} Thus a simple energy argument seemingly should favor the mixed chirality state. However, in the atomic basis we cannot draw the normal-state Fermi surface, and it is thus not clear what a gapped or gapless order parameter means for the quasiparticle energy, which is the important quantity. As a matter of fact, from the quasiparticle energy spectrum in Eq.~(\ref{eq:EQP}), we know that the $d \pm id'$ solutions have the same quasiparticle excitations in both valleys and thus the same gap structure on normal-state Fermi surfaces centered around $K$ and $K'$. Therefore, the mixed chirality state cannot be argued to appear in the $t$-$J$ model from a simple energy argument.

In the case of strong on-site repulsion and its associated $t$-$J$ model, the mixed chirality state is also problematic from a more fundamental physical point of view.
The $t$-$J$ model has very short range real-space pairing (on nearest-neighbor bonds), which corresponds to extended pairing in $\bfk$-space. For the honeycomb lattice close to half filling this dictates that the two electrons forming a Cooper pair belong to different Fermi surfaces and they form a zero-momentum paired state, i.e.~the two electrons have momenta $\bfk$ and $-\bfk$, respectively.
For zero-momentum pairing it is the symmetry of the order parameter over the {\it whole} Brillouin zone that is important, as discussed in Section~\ref{sec:symmetry}.
For the mixed chirality state, with $d+id'$ at $K$ and $d-id'$ at $K'=-K$, the $d$-wave part has even parity over the full Brillouin zone, whereas the $id'$-wave part has odd parity. However, a superconducting order parameter needs to be fermionic in nature, which means that even spatial parity leads to a spin-singlet state (for intraband pairing), while odd parity gives a spin-triplet state. Thus the mixed chirality state is necessarily a mixture of $d$-wave spin-singlet and $id'$ spin-triplet parts. This is not compatible with the superconducting state in the $t$-$J$ model which always have a spin-singlet nature, as clearly seen in e.g.~Eq.~(\ref{eq:HMF}). 
Moreover, if one still writes a mixed chirality solution and then tries to do an inverse Fourier transform back to real space in order to determine the bond  pairing correlations, only the $d$ part survives, whereas the $id'$ part results in zero contributions in all directions of the nearest-neighbor bonds. 
Based on this we conclude that the mixed chirality solution, with different chiralities in different valleys located at opposite momenta, is not a physically viable solution for models with zero-momentum spin-singlet pairing on the honeycomb lattice, which is the case in the $t$-$J$ model. It is perhaps in principle possible to get a mixed chirality phase for more complex models, for example with spin-orbit coupling or long-range interactions, but we here focus on the strong short-range electron-electron interaction limit of the simplest single-orbital spin-degenerate honeycomb lattice.

\begin{figure}[thb]
\includegraphics[scale = 0.23]{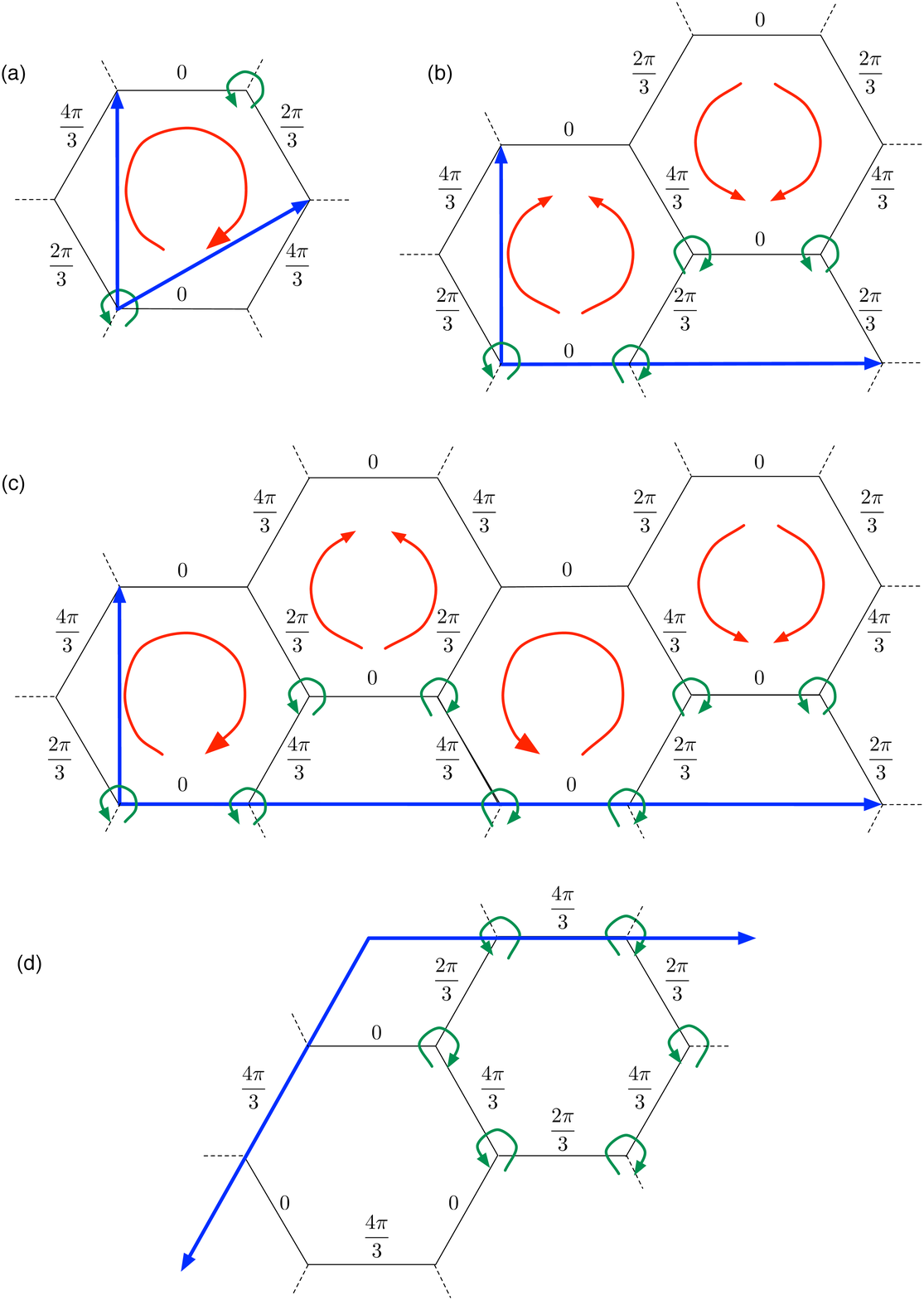}
\caption{\label{fig:cells} (Color online). Different sized unit cells on the honeycomb lattice indicated by the blue unit vectors and with the complex phase of the chiral $d$-wave bond order parameter displayed on each bond. Red (half) circular plaquette-centered arrows mark the winding (increasing angles) of the order parameter phase around each plaquette. Green circular site-centered arrows mark the winding of the order parameter phase around each site. Original two-site unit cell (a), four-site unit cell (b), and eight-site unit cell (c) with the phase of the order parameter on the horizontal bonds fixed to zero. (d) Six-site unit cell with a Kekul\'{e} distortion.
}
\end{figure}

The negative result above for the mixed chirality state begs the question of whether there is any other way to effectively cancel the chirality for the chiral $d$-wave superconducting state in the $t$-$J$ model? The $t$-$J$ model dictates nearest-neighbor bond pairing, which results in a zero-momentum paired state, with a spin-singlet spin structure. The only freedom left in the pure superconducting phase is the possibility of breaking translation symmetry by extending the real-space unit cell. In reciprocal space such real-space modulations lead to a reduction of the size of the first Brillouin zone. This could potentially help forming a state with mixed chirality or otherwise a zero net sum chirality, as it is the spatial symmetry consideration over the full Brillouin zone which is the underlying problem above. 
Thus, what we are looking for is a spin-singlet state with local chiral $d$-wave bond solution character, i.e., $\Delta \sim (1,\exp(\pm i2\pi/3),\exp(\pm i 4\pi/3))$ on each three nearest-neighbor bonds, but where we allow the unit cell to be larger than the original two-site cell, such that the same phase winding of the $d$-wave pattern is not repeated every two sites. This represents a mixing of chiralities in real space. Note that all these states are automatically orthogonal to the extended $s$-wave state.
In order to guide a search for a viable such state on larger unit cells we first study the $d+id'$-wave in the original two-site unit cell in Fig.~\ref{fig:cells}(a). For simplicity we fix the overall complex phase of the superconducting state such that the horizontal bond has a vanishing phase. We see that the order parameter phase winds twice around each plaquette (red plaquette arrow) creating a finite plaquette flux. Around each site the order parameter winds once, but it adds up to a $4\pi$ winding since both sites have the same winding direction. This is fully consistent with the winding number $\mathcal{N} = 2$ for the chiral $d+id'$ state. 
%
% FIGURE:
%
% 4 SITES:
\subsection{Four-site unit cell}
To find possible candidates for a state where the overall chirality is canceled in the unit cell, we start by studying the simplest expanded unit cell which contain four sites. Without loss of generality we again fix the phase on the leftmost horizontal bond to zero, and then assume a counterclockwise winding around the leftmost site. Assuming we can only put the phases $0$, $\frac{2\pi}{3}$, or $\frac{4\pi}{3}$ on each bond, we immediately arrive at the phase pattern in Fig.~\ref{fig:cells}(b) if we avoid repeating the two-site unit cell. Here the total site winding cancels over the four sites. The plaquette winding is also zero on each plaquette. This is thus a state where the overall chirality is zero. This state can be understood in terms of a system with vertical ferromagnetic order but antiferromagnetic horizontal order. Thus to recover the original state time-reversal and translation by one original two-site unit cell is needed in the horizontal direction.

We study this four-site unit cell using effective values $\tilde{t}$ and $\tilde{J}$, where we include the statistical weighting factors in the values of $\tilde{t}$ and $\tilde{J}$. We also include $\chi$ into the effective value of $\tilde{t}$, as it only renormalizes the band structure. These simplifications lead to the mean-field Hamiltonian:
%
% EQUATION:
\begin{align}
\label{eq:HMFtilde}
& H_{\tilde{t}\tilde{J}} = -\tilde{t} \sum_{\langle i,j\rangle,\sigma} a^\dagger_{i\sigma}b_{j\sigma} + {\rm H.c.}
+ \tilde{\mu} \sum_{i,\sigma} a^\dagger_{i\sigma}a_{i\sigma} + b^\dagger_{i\sigma}b_{i\sigma} \nonumber \\
& -\! \sum_{i, \alpha} \frac{\tilde{\Delta}_{i,\alpha}}{2} (a_{i\uparrow}^\dagger b_{i+\bfR_\alpha\downarrow}^\dagger \! - \! a_{i\downarrow}^\dagger b_{i+\bfR_\alpha\uparrow}^\dagger) + {\rm H.c.} +  \sum_{i,\alpha} \frac{|\tilde{\Delta}_{i,\alpha}|}{2J},
\end{align}
where $\tilde{\Delta}_{i,\alpha} = \tilde{J} \langle a_{i\downarrow} b_{j\uparrow} - a_{i\uparrow} b_{j\downarrow}\rangle$ is now the order parameter on bond $\alpha$ in the $i$th two-site cell. For example, for the four-site unit cell, we have different $\tilde{\Delta}_{i,\alpha}$ for $i$ and $i+1$, as illustrated in Fig.~\ref{fig:cells}(b).
The Hamiltonian in Eq.~(\ref{eq:HMFtilde}) allows for a straightforward solution with no formal constraint on the site occupancy, while still qualitatively capturing the relevant physics.

For a four-site real-space modulation the Brillouin zone in reciprocal space is half as big as the original Brillouin zone. This leads to the Dirac points, originally at the corners of the Brillouin zone at $K = 2\pi/a (0,2/3)$ and $K'=-K$ ($a$ is the length of the lattice unit vectors) now being ``folded'' into $K_2 = 2\pi/a (0,1/3)$ and $K'_2 = -K_2$. Performing a Fourier transform of the Hamiltonian in Eq.~(\ref{eq:HMFtilde}) to this reduced Brillouin zone we can write down the Hamiltonian in $\bfk$-space and then solve self-consistently for the size $\tilde{\Delta}_0$ of the order parameter. 
In order to find the most favorable state we compare their total free energy at zero $T$:
%
% EQUATION:
\begin{align}
\label{eq:FreeE}
F = \sum_{\bfk, E_{\bfk}<0}E_{\bfk} + \sum_{i,\alpha}\frac{\tilde{\Delta}_{i,\alpha}}{2J} - 2N\tilde{\mu},
\end{align}
where $E_{\bfk}$ are the quasiparticle energies of the Hamiltonian in Eq.~(\ref{eq:HMFtilde}) and $N$ is the total number of lattice sites.
More specifically, at filling fractions such that the normal-state Fermi surfaces are well-centered around the Dirac points, we compare the free energy of this four-site solution to that of the original two-site chiral $d+id'$-wave solution and also to that of the normal state. 
We find that the free energy of the four-site zero net flux solution is {\it  higher} than both the free energy of the original one-cell solution and the normal state. Thus the solution depicted in Fig.~\ref{fig:cells}(b) is not a physically viable superconducting state. 

% 8 SITES:
\subsection{Eight-site unit cell}
The result from the four-site unit cell shows that zero plaquette flux is not a stable configuration. It is also straightforward to show that it is not possible to have opposite  $4\pi$ windings on neighboring plaquettes. To achieve different plaquette windings it is necessary to insert  plaquettes with zero winding in-between. The smallest structure with plaquettes with opposite fluxes is therefore an eight-site unit cell, as displayed in Fig.~\ref{fig:cells}(c). Here the overall plaquette flux is zero since the overall winding cancels on the four plaquettes in the unit cell, even though some individual plaquettes now have finite fluxes. The total site phase winding also averages to zero.

The Brillouin zone for this eight-site real-space modulation is half that of the four-site unit cell. The Dirac points are still at $K_2 = 2\pi/a (0,1/3)$ and $K'_2 = -K_2$. We again use the Hamiltonian in Eq.~(\ref{eq:HMFtilde}), consider 
only systems with Fermi surfaces highly centered around the Dirac points, solve self-consistently for the order parameter, and compare the free energy of this eight-site state to that of the original two-site chiral $d+id'$-wave state and the normal state. We find that the free energy of eight-site solution is {\it higher} than both the free energy of the original one-cell solution and the normal state, which means that this is not a preferred state.
In fact, the structure in Fig.~\ref{fig:cells}(c) can be seen as two-site thick domains with different chiralities, with the zero flux two-site structures in-between forming minimum width domain walls between the two different chiral states. It is thus not surprising that this structure has a higher energy than solutions without any domain walls. This line of argument also extends to the four-site unit cell, which can be seen as only containing domain walls. Based on these results we conclude that even larger unit cells with alternating plaquette fluxes and/or zero-flux plaquettes will not provide a superconducting state with lower energy than the original, single chirality state.

% 6 SITES:
\subsection{Six-site unit cell}
The above results for the four- and eight-site unit cells shows that plaquettes with net zero flux are not energetically favorable and that they in fact can be viewed as domain walls. These unit cells represent the simplest real-space modulations of the chiral $d+id'$ state with the restriction that the phases on the horizontal bonds (or along another nearest-neighbor bond direction) are zero. This latter restriction is incidentally the same as in the (unphysical) mixed chirality $d \pm id'$ solution since the part that changes across the two different Fermi surfaces, i.e., $id'$, has a zero component on that bond. 
Here we finally relax also this restriction. The simplest real-space modulation with different phases on the horizontal bonds is a six-site unit cell, since the four-site unit cell requires the phase on the two horizontal bonds to be equal. The six-site unit cell allows for the so-called Kekul\'{e} distortion, which is a rather well-known distortion for the honeycomb lattice. It has recently been shown that it is possible to find two real-valued fully gapped spin-triplet Kekul\'{e} superconducting states if one starts with a sufficient strong attractive nearest-neighbor interaction and that these states can be energetically favorable.\cite{Roy10} 
The two states investigated had a $d_{x^2-y^2}$ and $d_{xy}$ Kekul\'{e} modulation.
We arrive at the same Kekul\'{e} real-space variation but for a spin-singlet order parameter as required by the $t$-$J$ model. In Fig.~\ref{fig:cells}(d) we show the complex chiral $d$-wave Kekul\'{e} modulation. There is for the Kekul\'{e} chiral $d$-wave structure no well defined winding per plaquette but the site phase winding averages out to zero. Thus this is a state which again presents a possibility of overall cancellation of the chirality.

In order to perform a straightforward study of the Kekul\'{e} pattern we again use the Hamiltonian in Eq.~(\ref{eq:HMFtilde}). At half filling in the original two-site unit cell an interaction strength of $\tilde{J}/\tilde{t} > 3.8$ is needed in order to reach the superconducting state. Also, at half filling the extended $s$- and the $d$-wave solutions are degenerate in the original two-site unit cell.\cite{Black-Schaffer07} It is only by doping away from half filling that the $d$-wave solutions are energetically favored (since $g_t = 0$ prevents superconductivity at half filling there is no problem with the statement that $d$-wave is always the favored symmetry in the $t$-$J$ model). 
Performing the exact same calculation to find the quantum critical point for the two different real valued Kekul\'{e} $d$-wave patterns, we find that both of these $d$-wave solutions require $\tilde{J}/\tilde{t} > 4.2$. Thus, both spin-singlet Kekul\'{e} $d$-wave patterns are notably less favorable than the extended $s$-wave solution at zero and low doping levels. Since the Kekule\'{e} chiral $d$-wave solution is only a complex combination of these two solutions with the same node at the Dirac points, it has the same quantum critical point at $\tilde{J}/\tilde{t} = 4.2$. We therefore conclude that close to half filling any Kekul\' e spin-singlet $d$-wave structure is less favorable than the extended $s$-wave solution, and again this is not a competitive candidate for a zero-chirality $d$-wave state. 
This result can be understood by considering that the Kekul\'{e} distortion folds both the $K$ and $K'$ points onto the zone center at $\Gamma$. At $\Gamma$ the $d\pm id$-wave solutions are both gapless and only the extended $s$-wave solution is gapped, which should favor the $s$-wave solution. Note that this is different from the situation at the $K,K'$ points where both the extended $s$-wave and the chiral $d+id'$ states are gapless. Therefore, there is in the original two-site unit cell no such simple energy argument as to favor one solution over the other. Instead, the $d$-wave solutions ultimately wins in the $t$-$J$ model in the original two-site cell since they are orthogonal to on-site pairing which is disfavored in systems with strong electron-electron repulsion.

\subsection{Larger unit cells}
To conclude the search for a mixed or zero net sum chirality state we note that, while an arbitrary large extended unit cell leads to a reduced Brillouin zone, the Dirac cones are still folded onto opposite momenta, $K_i$ and $-K_i$. 
As long as this momentum is non zero, a solution with different chiralities in different valleys necessarily has the $id'$ part in a spin-triplet configuration, which is prohibited in the $t$-$J$ model, or any other model with spin-singlet superconductivity. Both the four- and eight-site solutions discussed above also show that real-space realizations producing allowed zero net sum chirality states do not produce a favorable superconducting state for $K_i$ non zero.
Only if both Dirac cones are folded onto the zone center at $\Gamma$ does the spin-triplet requirement of the $id'$ part go away. However, in this case, the two Dirac cones become fully overlapping and they then likely hybridize, preventing different phases on the two Fermi surfaces. The Kekul\'{e} distortion discussed above is one zero net sum chirality example with $K_i = 0$ where hybridization causes a mass gap to open in the Dirac spectrum.\cite{Ryu09} While the Kekul\'{e} spin-singlet chiral $d$-wave real-space modulation is physically feasible, it is energetically unfavorable compared to the single chirality $d$-wave state.

% -------------------------------------------------- %
% CONCLUSIONS:
% -------------------------------------------------- %
\section{Discussion and conclusions}
\label{sec:conclusions}
The honeycomb lattice close to half filling offers unique opportunities to host exotic superconducting states driven by electron-electron interactions. Due to the sixfold lattice symmetry and disjoint Fermi surfaces at $K,K'$ several possible order parameters produce a fully gapped quasiparticle spectrum even though they contain nodal lines in the Brillouin zone, as displayed in Table~\ref{tab:sym}.
This includes one of the possible spin-triplet $f$-wave states, which has a fully gapped $s_{\pm}$-wave symmetry close to $K,K'$ and has been found to be the favored state for the Hubbard model in the weak-coupling limit. \cite{Raghu10, Nandkishore14}
In the limit of strong on-site pairing, as described by the $t$-$J$ model, the pairing instead has a spin-singlet configuration due to the antiferromagnetic Mott state in the undoped limit. The sixfold lattice symmetry then gives rise to the time-reversal breaking chiral $d+id'$-wave state, which is also a fully gapped state.\cite{Black-Schaffer07, Honerkamp08, Pathak10, Ma11, Wu13, Gu13} We have here provided QMC results showing that the chiral $d+id'$-wave is the favored superconducting symmetry over a wide range of $U$ values spanning the intermediate to strong coupling regimes.
Only at very large doping levels beyond the van Hove point is an extended $s$-wave scenario plausible. \cite{Wu13, Lothman14}

Most importantly, by investigating the quasiparticle energy spectrum and the symmetry aspects of the chiral $d$-wave state, we have also studied the feasibility to find a recently proposed mixed chirality $d$-wave state,\cite{Tran11,Wu13} where the pairing is $d+id'$ in one valley and $d-id'$ in the other valley.
However, due to zero-momentum pairing and a spin-singlet configuration in the $t$-$J$ model, we find that a mixed chirality state is not possible to achieve without reducing the translational symmetry.
To extend the search for a mixed or other zero net sum chirality states we investigated extended unit cells with an overall zero phase winding (flux) of the complex $d$-wave order parameter. Extended cells with alternating or zero plaquette fluxes can be seen as different domain and domain walls configurations, respectively, and we find that such states are not energetically favorable. We also find that the Kekul\'{e} chiral $d$-wave distortion, which produces a state with no definable plaquette flux, do not give a superconducting state with lower energy than the single chirality $d$-wave state. We thus find no favored mixed or otherwise zero net sum chirality states using simple real-space modulations. 
Moreover, any real-space modulation with the two Dirac points at non zero $K_i,-K_i$ cannot produce a mixed chirality state due to the spin-singlet constraint on the order parameter. Only when $K_i = 0$ is this restriction lifted but then any valley hybridization leads to the same phase of the order parameters in the two valleys.
Therefore we conclude that the single chirality $d+id'$-wave state is likely the most stable state even in the presence of disjoint Fermi surfaces on the lightly doped honeycomb lattice in the limit of strong electron-electron interactions as described within the $t$-$J$ model. It is, however, possible that including spin-orbit coupling\cite{Hyart12, You12, Scherer14} or longer-range Coulomb interactions, might lead to other exotic states, but that goes beyond the scope of this work.

% -------------------------------------------------- %
% ACKNOWLEDGMENTS
% -------------------------------------------------- %
\begin{acknowledgments}
We thank A.-M.~Tremblay, B.~Dou\c{c}ot, and T. Liu for discussions related to this work. We have also benefited from discussions at CIFAR meetings in Canada and at the 2013 Carg\`{e}se summer school on topological phases in condensed matter and cold atom systems.  
This work was supported by the Swedish Research Council (VR) and the G\"{o}ran Gustafsson foundation (A.B.-S.), the Natural Sciences and Engineering Research Council of Canada (NSERC) (W.W.), and by the Labex Palm at Paris-Saclay (K.L.H.).
\end{acknowledgments}

% -------------------------------------------------- %
% BIBLIOGRAPHY:
% -------------------------------------------------- %
%\bibliographystyle{apsrevmy}
%\bibliography{DiracMaterials}

\end{document}